\documentclass[prl,twocolumn,letterpaper,superscriptaddress,amsmath,amssymb,showpacs,longbibliography]{revtex4-1}

\usepackage{float}

\usepackage{subfigure}
\usepackage{graphicx}
\usepackage{bm,mathrsfs}
\usepackage{braket}
\usepackage{amsmath}
\usepackage{amssymb}
\usepackage{esint}
\usepackage{setspace}
\usepackage{hyperref}
\usepackage[table]{xcolor}
\usepackage{colortbl}
\usepackage{mathtools}
\usepackage{multirow}
\usepackage{bbold}
\usepackage{pbox}
\usepackage{ulem}

\definecolor{gray1}{gray}{0.9}
\definecolor{gray2}{gray}{0.6}
\definecolor{gray3}{gray}{0.3}

\makeatletter

\begin{document}

\title{Doping-limitations of cubic boron nitride: effects of unintentional defects on shallow doping}

\author{Tamanna Joshi} 
\affiliation{Department of Physics and Astronomy, Howard University, Washington, D.C. 20059, USA}
\author{Pankaj Kumar}
\affiliation{Department of Physics and Astronomy, Howard University, Washington, D.C. 20059, USA}
\author{Bipul Poudyal}
\affiliation{Department of Physics and Astronomy, Howard University, Washington, D.C. 20059, USA}
\author{Sean Paul Russell}
\altaffiliation{Current address: Department of Mathematics, Indiana University, Bloomington, IN, USA}
\affiliation{Department of Physics, University of Evansville, Evansville, IN, USA}
\author{Priyanka Manchanda}
\affiliation{Department of Physics and Astronomy, Howard University, Washington, D.C. 20059, USA}
\author{Pratibha Dev}
 \email{pratibha.dev@howard.edu}
\affiliation{Department of Physics and Astronomy, Howard University, Washington, D.C. 20059, USA}

\begin{abstract}
Cubic boron nitride (cBN) is an ultra-wide bandgap, super-hard material with potential for extreme-temperature and -pressure applications.  A proof-of-principle p-n junction using cBN was demonstrated almost three decades ago. However, to date, there remain two unresolved challenges that prevent its practical use  in technologies: (i) it is difficult to produce high-quality cBN films and (ii) it is difficult to controllably n- and p-dope its matrix. In this theoretical work, we study the reasons for doping-limitations, which is an acute issue in realizing cBN-based electronics. In particular, we find that different unintentionally-present intrinsic and extrinsic defects act as compensating defects and/or introduce trap states. In turn, the presence of these defects and their complexes affect the incorporation, as well as the electronic structure properties, of shallow dopants [silicon and beryllium], which are introduced intentionally to n- and p-dope cBN. Our analysis of doping-limitations provides a path towards finding solutions for controllably n- and p-doping cBN.
\end{abstract}


\maketitle

\section{I. Introduction }
Among group III-nitrides, cubic boron nitride (cBN) is considered a potential candidate for  high power and high temperature electronic devices~\cite{chen2020ultrahigh, Natalia_Review_2017, chen2015misfit, bello2005cubic, kirschman1999cubic,  mishima1987,mishima1988} due to its fascinating mechanical and electronic properties. These properties include: an ultra-wide band gap of about 6.1\,eV~\cite{miyata1989-bandgap}, exceptional hardness, high thermal conductivity~\cite{Haubner2002-ref}, and corrosion resistance, which are useful for any device to operate in harsh environment. 

The key to using cBN for high performance electronic applications is to control its electrical conductivity. In turn, this requires the development of methods that controllably n- and p-dope cBN, creating ionizable delocalized (shallow) impurity states. Over the past several decades, different experiments have studied intentional n- and p-doping of cBN~\cite{wentorf1962preparation,mishima1987,mishima1988,shishonok2016xrd,li2014,ying2011, he2008,nose2006electric,shishonok2005luminescence,taniguchi2002appearance,Haque_2021}. The n-type dopants investigated include: silicon (Si)~\cite{mishima1987,li2014,ying2011}, sulphur (S)~\cite{taniguchi2002appearance} and carbon (C) ~\cite{Haque_2021}; p-type dopants explored include: beryllium (Be)~\cite{mishima1987,he2008}, magnesium (Mg)~\cite{Kojima_2009} and zinc (Zn)~\cite{nose2006electric}.
In spite of these efforts, the research to create cBN-based electronics is still in early stages of development, and has been hampered by two persistent challenges: (i) difficulties in synthesizing high quality pure-phase cBN crystals, (ii) formation of different unintentional defects during cBN growth, which makes it hard to controllably n- and p-dope cBN. Depending on growth conditions, cBN not only hosts intrinsic defects (e.g. nitrogen vacancies)~\cite{fanciulli1993}, but also unintentional extrinsic defects~\cite{bello2005cubic,taniguchi2002appearance,mishima1988,wentorf1962preparation}. Carbon and oxygen (O) are the most common inadvertently-introduced impurities. For example, a significant amount of such impurities are typically detected by X-ray photoelectron spectroscopy of as-grown boron nitride~\cite{Widmayer-CandO-1999}. In fact, the characteristic amber color of cBN is believed to be related to a high concentration of oxygen-related defects~\cite{taniguchi2001}. Some of these unintentional defects introduce deep states within the band structure. The inefficient performance of p-n junctions made from cBN is associated with these localized deep states that act as charge traps~\cite{mishima1987}. 

On the theoretical side, many studies have investigated properties of different intentionally- and unintentionally-introduced defects in cBN~\cite{Gubanov1996, Gubanov-1997, Orellana1999, Orellana2000, Dev2008,Tesfaye2014,li2014-dft, mcdougall2017theoretical,weston2017hole, li2018first}. Nevertheless, how the presence of different unwanted defects affects the properties of the shallow dopants has not been investigated and this important issue remains unresolved. In this theoretical work, we carried out a systematic study of the effects of unintentional impurities on the characteristics of doped cBN, uncovering the reasons for doping-limitations, a bottleneck in realizing cBN-based electronics. In particular, we investigated C- and O-related defects, due to their abundance in as-grown cBN. We find that, from a thermodynamical point of view, substitutional oxygen, and two defect complexes (carbon- and oxygen-based) can form in large numbers. We show that the presence of such unintentional defects adversely affects intended doping of cBN, using two shallow dopants (Si and Be) in our proof-of-principle study.  This analysis provides an important step towards understanding the influences of unwanted defects and how these can be minimized for controllably n- or p-doping cBN. 

%



%
%

\begin{figure*}
	\centering
	\includegraphics[width=0.95\linewidth]{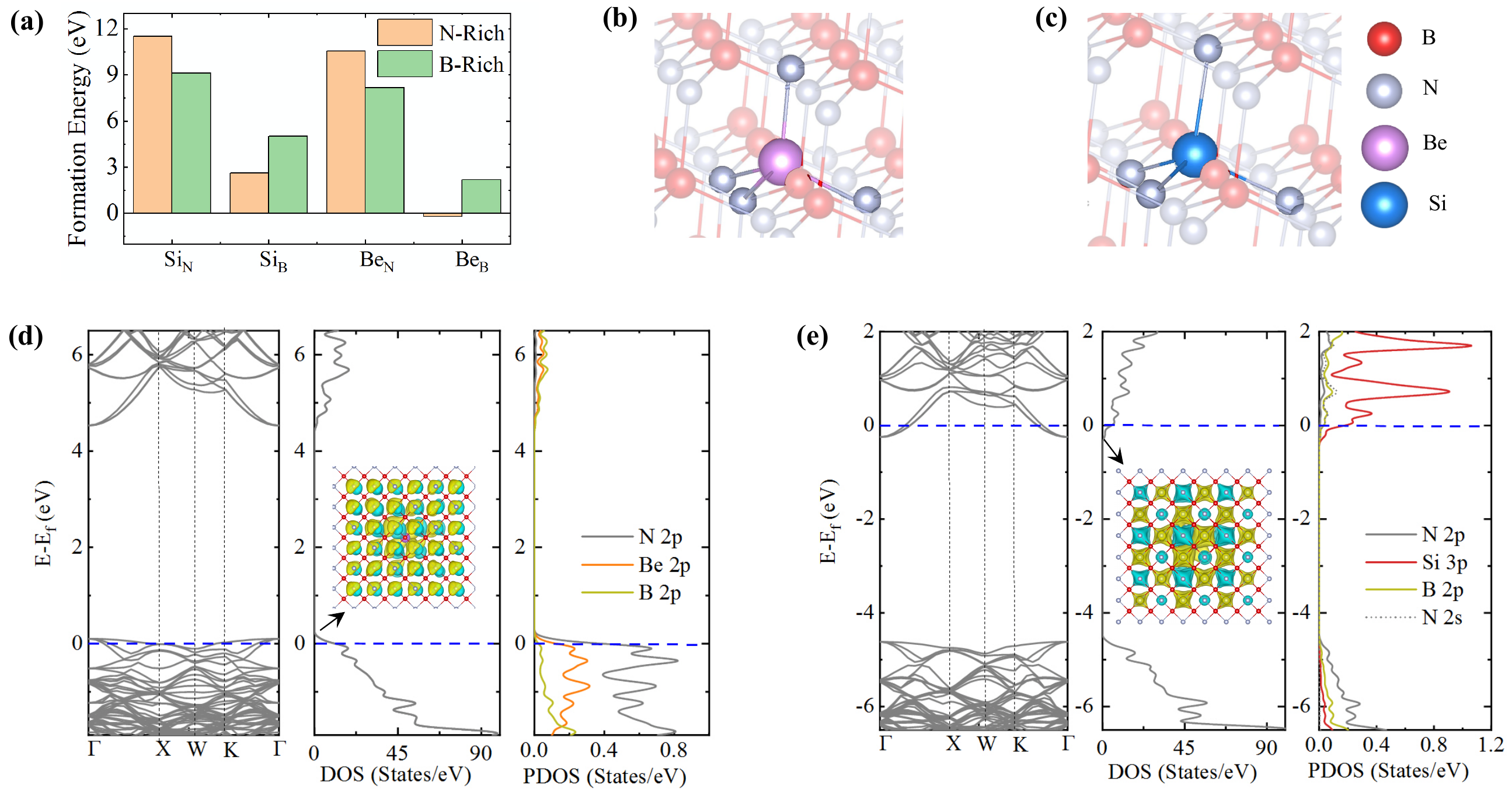}
	\caption{Doping cBN: (a) Formation energies for intentional shallow dopants (Be and Si) in N-rich and B-rich conditions, indicating that B-site substitution by either of the dopants is more favorable than the N-site substitution. Substitutional defect, (b) $\mathrm{Be_{B}}$ for p-doping cBN, and (c) $\mathrm{Si_{B}}$, for n-doping cBN. Band structures, total density of states (DOS) and projected density of states (PDOS) for (d) $\mathrm{Be_{B}}$ and (e) $\mathrm{Si_{B}}$. The zero of the energy is set at the Fermi energy (E$ _\text{{f}}$). The inset in (d) shows the shallow acceptor state (at the $\Gamma$-point) introduced just above the Fermi level, with most significant contributions from the $2p$-derived orbitals of Be and N-atoms. The inset in (e) shows the shallow donor state (at the $\Gamma$-point) introduced just below the Fermi level, with contributions from nearly all atoms, including Si $3p$. Yellow (blue) colors correspond to positive (negative) isovalues. }
	\label{fig:Shallow_Dopants}
\end{figure*}

\section{Computational Details}
Spin-polarized calculations based on density functional theory (DFT) were performed with Quantum Espresso package~\cite{QE-2009, QE-2017}. The generalized gradient approximation (GGA)~\cite{GGA} of Perdew-Burke-Erzerhof (PBE)~\cite{PBE} was used to approximate the exchange-correlation functional. As ultrasoft pseudopotentials~\cite{vanderbilt1990soft} were used, we found that a plane-wave energy cutoff of 65\,Ry was sufficient to give converged results. The optimized lattice constant of a cBN conventional unit cell was found to be 3.62\,\AA{}, consistent with other theoretical ~\cite{Gubanov1996,Orellana2000} and experimental~\cite{Haubner2002-ref} reports.  A 3$\times$3$\times$3 supercell consisting of 216 atoms was used to study the effect of intentionally-introduced and unintentionally-present defects. We used a $\Gamma$-centered, 2$\times$2$\times$2 k-point grid to sample the Brillouin zone of the supercell. Forces in every structure were fully relaxed until they were reduced below 10$^{-4}$\,Ry/a.u.

In order to quantify the ease with which these defects are formed, we calculated their formation energies. The formation energy, $\Delta E_{form}[X]$, of a defect [$X$] was calculated using:
\begin{equation}
    \Delta E_{form}[X]=E_t[X]-E_t[bulk]-\sum_i n_i \mu_i
    \label{eq:eq1}
\end{equation}
Here $E_t[X]$ is the total energy of the defective system, $E_t[bulk]$ is the total energy of the pristine/defect-free system, $n_i$ represents the number of  atoms of species $i$ that have been added or removed, and $\mu_i$ is the corresponding chemical potential. Chemical potentials vary depending on different experimental conditions under which defects are formed. For cBN, the chemical potentials are constrained by: $\mu_\text{B} + \mu_\text{N} = \mu_\text{cBN}$. In the boron-rich (B-rich) condition, we used the upper bound for boron's chemical potential: $\mu_\text{B} = \mu_\text{B(bulk)}$, where $\alpha$-boron was used to obtain $\mu_\text{B(bulk)}$, and $\mu_\text{N}$ was obtained as $\mu_\text{N} = \mu_\text{cBN}-\mu_\text{B}$. Similar considerations are taken into account for the nitrogen-rich (N-rich) condition, in which case nitrogen gas (N$_2$) was used to obtain $\mu_\text{N}$. In addition, graphene was used as a reference for carbon impurities and oxygen gas for oxygen impurities.

\section{Results and Discussion}
In this work, we have studied different properties of intentionally-doped cBN in the presence of inadvertently-present defects, elucidating how the presence of the unwanted defects can adversely influence n- and p-doping of cBN. Section I reports the electronic structure properties of cBN when doped with p-type (Be) and n-type (Si) dopants. In Section II, we report the electronic structure properties of unintentional defects and their complexes. Section III details the effects of these unintentional defects on the structural and electronic properties of the shallow dopants. 


\subsection{I. Shallow Dopants -- Be and Si -- in \lowercase{c}BN}

In cBN, beryllium~\cite{li2018first, he2008,Gubanov-1997} and silicon~\cite{li2014,ying2011,Gubanov-1997} have been successfully used as acceptor and donor impurities, respectively.  Figure~\ref{fig:Shallow_Dopants} (a) shows the formation energies of four substitutional defects -- beryllium at a boron site (Be$_\text{B}$), beryllium at a nitrogen site (Be$_\text{N}$), silicon at a boron site (Si$_\text{B}$) and silicon at a nitrogen site (Si$_\text{N}$) -- in both B-rich and N-rich conditions.  Under any condition, lower formation energies for the B-site substitutional defects imply that the shallow dopants will preferentially go into a B-site as compared to an N-site. In particular, whether under B-rich or N-rich conditions, the energy of forming $\mathrm{Si_{B}}$ is much smaller than that for $\mathrm{Si_{N}}$, even though a silicon-atom is amphoteric and can, in principle, substitute both the cation and the anion in cBN. Similarly, while both defects --  $\mathrm{Be_{B}}$  and $\mathrm{Be_{N}}$ -- should lead to a p-type conductivity, we find that only the former is energetically favorable. This can be traced back to the covalent radii of the atoms involved: Be (0.96\,\AA{}), Si (1.11\,\AA{}), B (0.84\,\AA{}), and N (0.71\,\AA{}). The larger differences in atomic radii of the dopant atoms and the N-atom, results in a greater distortion when a Be-atom or a Si-atom substitutionally replaces a nitrogen as compared to when they replace a boron atom.  Since Si- and Be-atoms are more likely to substitute at the B-sites, in the rest of the work, we have considered only $\mathrm{Be_{B}}$ and $\mathrm{Si_{B}}$ substitutional defects. These defects are shown in Fig.~\ref{fig:Shallow_Dopants} (b)-(c) In addition, Fig.~\ref{fig:Shallow_Dopants}(a) shows that under any condition, incorporation of Be at a B-site is energetically more favorable than Si at a B-site. This is because Be, with a covalent radius that is similar to boron, causes less distortion as compared to the silicon-atom substituting boron.  This might be one of the contributing factors that results in the challenges encountered in uniformly doping cBN with Si, which tends to segregate at the surface~\cite{Yin_Si_cBN_2008,Yin_Si_cBN_2014}.

Figure~\ref{fig:Shallow_Dopants}\,(d) shows the band structure, as well as total and projected densities of states (labelled as DOS and PDOS, respectively) of cBN with a beryllium substitutional ($\mathrm{Be_{B}}$). The Fermi level (E$_\text{f}$) is used as the reference energy.  The band structure shows that a highly dispersive, shallow band is introduced at the valence band edge.  Hence, the substitution of a trivalent boron atom with a divalent beryllium atom in cBN introduces a shallow acceptor (triplet) state, as can also be seen in the DOS plot. The PDOS plot shows the contributions to this shallow state from the beryllium's $2p$-orbital along with the $2p$-derived orbitals of one of the nearest neighboring nitrogen and a next-to-nearest neighboring boron atom. The inset in Fig.~\ref{fig:Shallow_Dopants}\,(d) is a plot of this shallow donor state in real-space, showing that this state is very delocalized (dispersive), with contributions from nearly all N $2p$-derived orbitals, as well as from Be $2p$-orbitals. The band structure and DOS plot in Fig.~\ref{fig:Shallow_Dopants}\,(e) for cBN doped with silicon (as $\mathrm{Si_{B}}$) show that the structure is n-doped when a tetravalent silicon atom replaces a trivalent boron atom. This substitution results in a shallow, highly-dispersive energy band just below the conduction band edge. The inset in Fig.~\ref{fig:Shallow_Dopants}\,(e) is a charge density plot corresponding to the defect-induced dispersive donor state (at the $\Gamma$-point) just below the Fermi level, showing the extended nature of the defect state in real space. We, therefore, find that the beryllium (as $\mathrm{Be_{B}}$) and silicon (as $\mathrm{Si_{B}}$) are p- and n-type dopants, respectively, which is in agreement with experiments~\cite{he2008, li2014}. Next, we consider different unwanted defects that are expected to be present in cBN.  These unwanted defects can potentially affect the incorporation of Si and Be at the B-site, as well as the change the properties of the doped cBN.

\begin{figure*}
    \centering
    \includegraphics[width=0.95\linewidth]{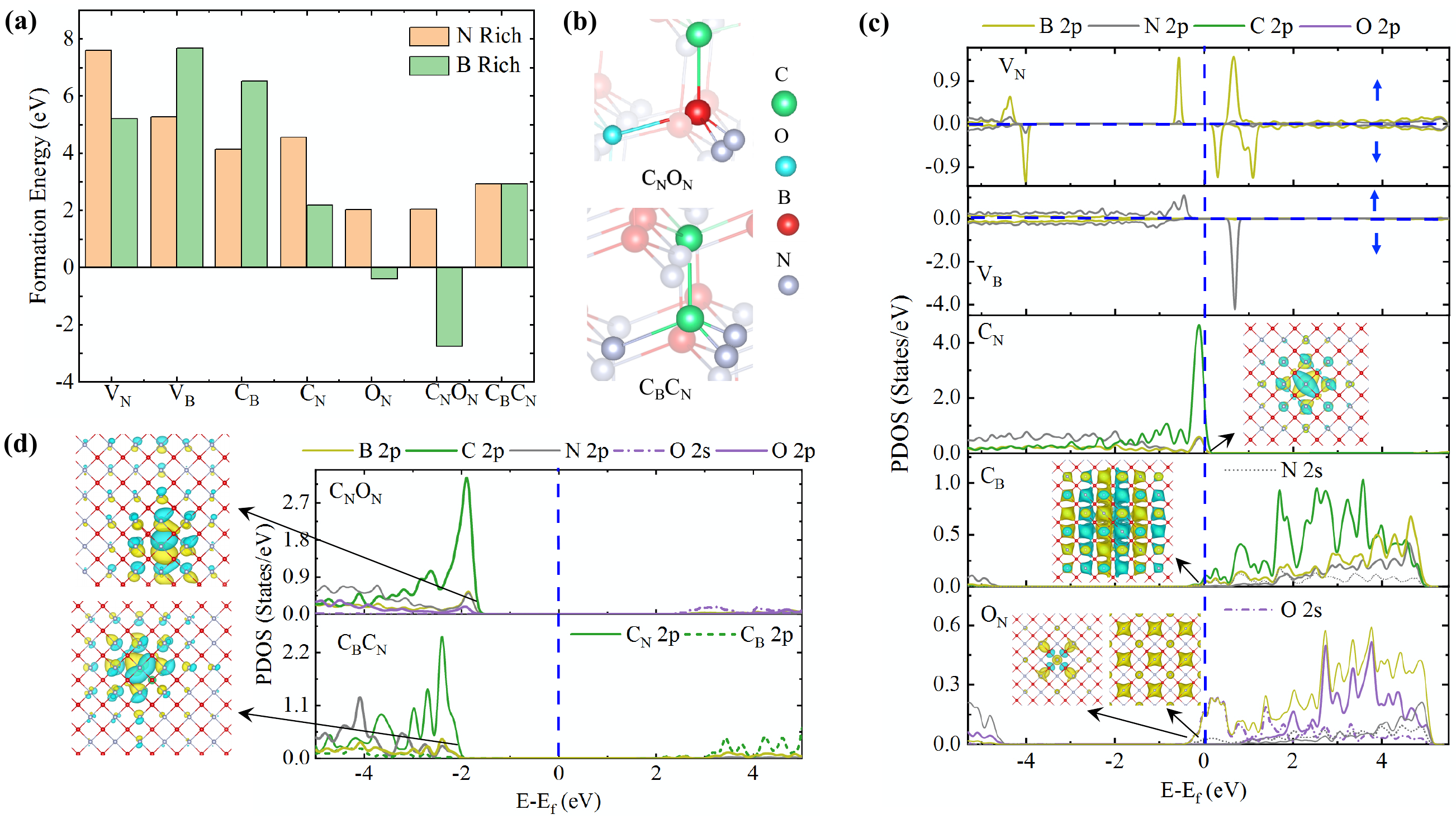}
    \caption{Unintentional defects: (a) Formation energies for different unwanted defects in N- and B-rich conditions. Negative formation energies of defects under different conditions imply that they are highly likely to be present.  (b) Structures of defect complexes -- C$_\text{{N}}$O$_\text{{N}}$ (top) and C$_\text{{B}}$C$_\text{{N}}$ (bottom) -- studied in this work. (c) PDOS plots for cBN with simple point defects. In the case of spin-polarized structures (V$_\text{N}$ and V$_\text{B}$), the DOS in the two spin channels is indicated by the blue arrows. The insets in the PDOS plots for C$_\text{N}$ and C$_\text{B}$ are the charge density plots for the defect states (at the $\Gamma$-point) introduced just above and below the Fermi level, respectively.  The insets in the PDOS plot for O$_\text{N}$ show the localized state, along with one of the two nearly-degenerate dispersive states introduced below the Fermi level by the defect. (d) PDOS for defect complexes, showing their insulating nature as the constituent point defects compensate each other in both defect complexes. The charge density plots show the localized nature of the defect-induced states at the valence band edge for both defect-complexes. The Fermi level is used as the reference energy in all PDOS plots. Yellow (blue) colors correspond to positive (negative) isovalues in all charge density plots.}
    \label{fig:unwanted_defects}
\end{figure*}

\subsection{II. Unintentional Defects in  \lowercase{c}BN}
While it is possible to both n- and p-dope cBN, one of the reasons cBN-based devices have not reached technological maturity is the large number of unwanted defects created during growth~\cite{wentorf1962preparation,mishima1987,mishima1988,Widmayer-CandO-1999,taniguchi2002appearance,bello2005cubic,mcdougall2017theoretical}. Figure~\ref{fig:unwanted_defects}\,(a) shows the formation energies of several unintentional intrinsic and extrinsic defects that are expected to be present in cBN. The native defects (V$_\text{B}$ and V$_\text{N}$) have much higher formation energies as compared to the extrinsic, simple point defects.  Carbon, which is amphoteric, can replace boron as well as nitrogen. We find that in boron-rich conditions, C$_\text{N}$ (p-type defect) is more likely to form than C$_\text{B}$ (n-type defect). On the other hand, the formation energies of the two defects are similar in nitrogen-rich conditions, implying that both defects -- C$_\text{N}$ and C$_\text{B}$ -- are equally likely to form. O$_\text{{N}}$ has a small formation energy in a N-rich environment and a negative formation energy of -0.38\,eV in a B-rich environment, suggesting an abundance of oxygen-related defects, in agreement with the experimental reports~\cite{Widmayer-CandO-1999,taniguchi2001}.  We also considered two defect-complexes formed from the point defects - C$_{\textrm B}$C$_{\textrm N}$ and C$_{\textrm N}$O$_{\textrm N}$ -- both of which are likely to form. In ascending order, the formation energies for unwanted defects and possible defect complexes are:

\noindent N-rich:  O$_\text{{N}}$  $<$ C$_\text{{N}}$O$_\text{{N}}$ $<$ C$_\text{{B}}$C$_\text{{N}}$  $<$  C$_\text{{B}}$  $<$  C$_\text{{N}}$  $<$  V$_\text{{B}}$  $<$  V$_\text{{N}}$

\noindent B-rich:  C$_\text{{N}}$O$_\text{{N}}$  $<$  O$_\text{{N}}$  $<$  C$_\text{{N}}$  $<$  C$_\text{{B}}$C$_\text{{N}}$  $<$  V$_\text{{N}}$  $<$   C$_\text{{B}}$  $<$  V$_\text{{B}}$



Out of the two defect complexes, C$_{\textrm N}$O$_{\textrm N}$ has a lower formation energy under any condition: $-2.75$\,eV and 2.04\,eV in B-rich and N-rich environments, respectively. The negative formation energy of this defect complex under B-rich conditions indicates that it is preferred, as compared to the isolated constituent substitutionals. The favorable formation of this defect is partly due to the electrostatic attraction between the donor-like defect, O$_\text{{N}}$, and the acceptor-like defect, C$_\text{{N}}$. For the same reason, in N-rich conditions, a C$_\text{{B}}$C$_\text{{N}}$ complex would form more readily than individual constituent point defects. The structures of the two defect complexes -- C$_{\textrm N}$O$_{\textrm N}$ and C$_\text{{B}}$C$_\text{{N}}$ -- are shown in Fig.~\ref{fig:unwanted_defects}(b).

Figures~\ref{fig:unwanted_defects}(c) and (d) show the PDOS of different point defects and their complexes. Both V$_\text{{N}}$ and V$_\text{{B}}$ introduce deep defect states, which are derived from the localized $2p$-orbitals of the boron and nitrogen atoms, respectively, surrounding the defects. In the presence of both V$_\text{{N}}$ and V$_\text{{B}}$, the structure becomes spin polarized, with former developing a net magnetic moment of $1\,\mu_{B}$ and $3\,\mu_{B}$, respectively. These results are consistent with earlier works on different polytypes of BN~\cite{Dev2008,Dev2020}. As the native vacancies have high formation energies, we have not considered them further in this work. Fig.~\ref{fig:unwanted_defects}(c) shows that the carbon impurity, incorporated as C$_\text{{N}}$, is an acceptor-like defect. C$_\text{{N}}$ introduces carbon's $2p$-derived deep states close to the valence band edge. The charge density plot of the defect state just above the Fermi level is plotted in the inset and shows the spatially localized nature of the state. On the other hand, the carbon impurity that is incorporated as C$_\text{{B}}$ is a shallow donor impurity. This can be seen in the inset, showing the charge density plot for the state just below the Fermi level. The PDOS plot for O$_\text{{N}}$ [Fig.~\ref{fig:unwanted_defects}(c)] indicates that boron's $2p$ and oxygen's $2s$-derived states contribute to the states below Fermi level. The insets in the PDOS plot for O$_\text{N}$ show the localized state, along with one of the two nearly-degenerate dispersive states introduced below the Fermi level by the defect [see Supplementary Note I for additional details].  It should be pointed out that amongst different simple point defects, C$_\text{{B}}$ is an ionizable impurity, which can by itself be used to n-dope cBN. This result is consistent with a recent experimental work, which successfully showed n-doping of cBN by carbon~\cite{Haque_2021}. However, the exact nature of the carbon-based defect within that experimental study remains unknown and requires further investigation.

The PDOS plots in Fig.~\ref{fig:unwanted_defects}(d) for the C$_{\textrm B}$C$_{\textrm N}$- and C$_{\textrm N}$O$_{\textrm N}$-defect complexes show the insulating nature of these structures. This is due to the compensation of charges in the complexes since each of them combines a p-type defect with an n-type defect. In addition, the C$_{\textrm N}$O$_{\textrm N}$ and C$_\text{{B}}$C$_\text{{N}}$ complexes show carbon $2p$-derived localized states at the valence band edge, reminiscent of the simple defect C$_\text{{N}}$, which is one of the constituent defects within the complexes. This can also be seen in the charge density plots of the defect states at the valence band edge, showing their spatial localization. The presence of such spatially-localized defect states are expected to be detrimental to shallow doping of cBN and its use in electronics. In order to explore whether these defects are responsible for the doping-bottleneck, we further studied properties of Si- and Be-doped cBN in the presence of these unwanted defects.

\begin{figure*}
    \centering
    \includegraphics[width=0.95\linewidth]{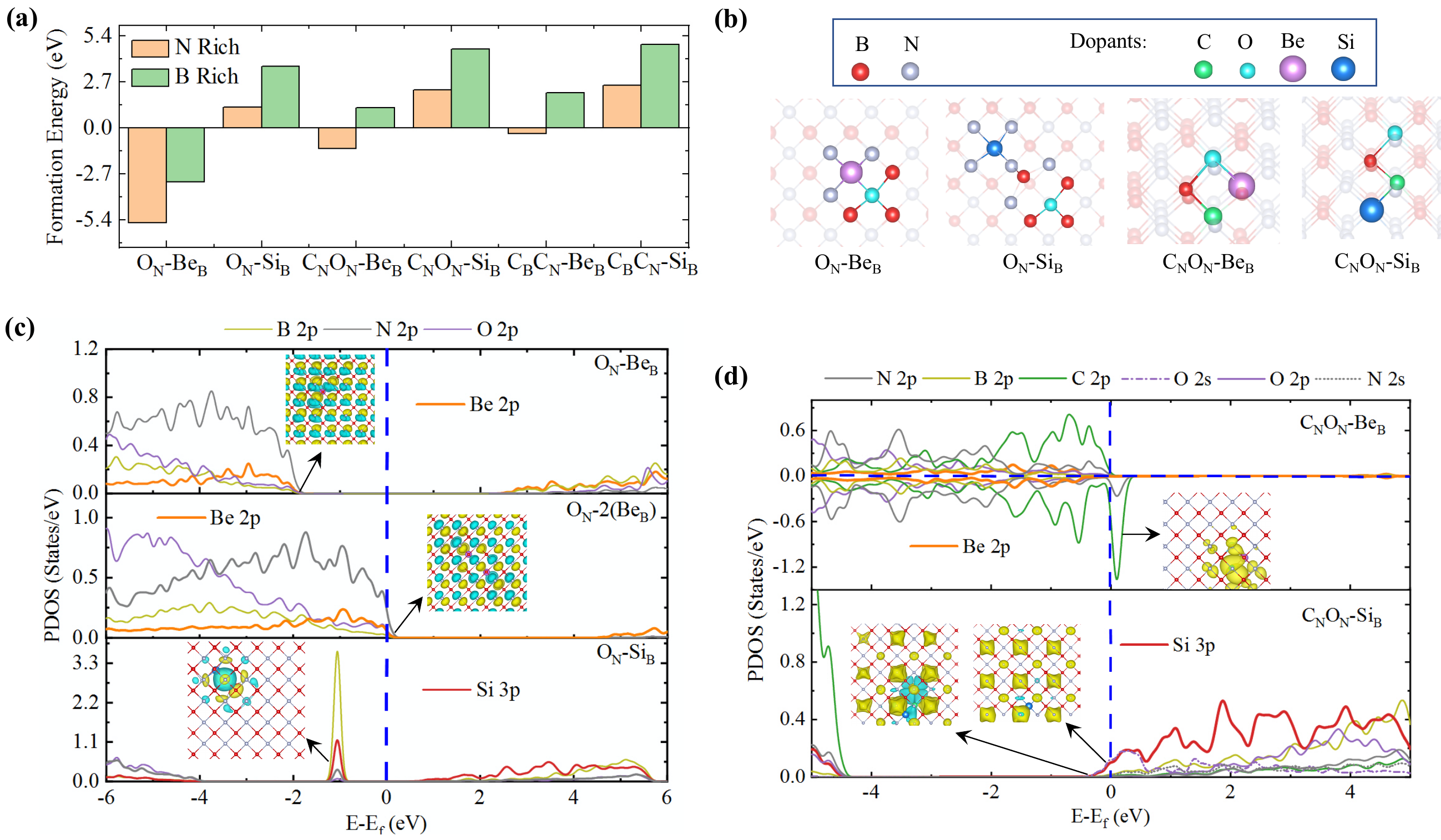}
    \caption {Shallow doping in presence of unintentional defects: (a) Formation energies for creating $\mathrm{Be_{B}}$ and $\mathrm{Si_{B}}$ in presence of unwanted defects in both N and B-rich conditions. In general, incorporation of Be is easier in N-rich conditions. This is especially the case in presence of oxygen-containing defects. (b) Lowest-energy optimized structures for $\mathrm{Be_{B}}$ and $\mathrm{Si_{B}}$ in presence of O$_\text{{N}}$ and C$_\text{{N}}$O$_\text{{N}}$. PDOS showing modification in electronic structure properties of cBN upon doping with Be and Si in presence of  (c) O$_\text{{N}}$, and (d) C$_\text{{N}}$O$_\text{{N}}$.  The inset in the top panel of (c) shows the isosurface plot of a shallow state (at $\Gamma$-point) at the valence band edge for cBN doped with equal concentrations of O$_\text{{N}}$ and $\mathrm{Be_{B}}$. The inset in the middle panel of (c) shows the state just above Fermi level for cBN doped with an additional $\mathrm{Be_{B}}$ (i.e. for concentration ratios $\mathrm{Be _{B}:O_{N}}::2:1$), while inset in the bottom panel in (c) shows the defect-induced polaronic state below Fermi level for $\mathrm{Si_{B}}$ in the presence of O$_\text{{N}}$. The inset in top panel of (d) [$\mathrm{Be_{B}}$ in presence of C$_\text{{N}}$O$_\text{{N}}$] is a charge density plot for the spatially-localized defect state just above the Fermi level, while the inset in the bottom panel [$\mathrm{Si_{B}}$ in presence of C$_\text{{N}}$O$_\text{{N}}$] is the real-space plot of the defect-induced delocalized state just below Fermi level. Yellow (blue) colors correspond to positive (negative) isovalues in all charge density plots.
}
    \label{fig:panel3-Shallow-doping-with-unintentional-impurities}
\end{figure*}

\subsection{III. Intentional Shallow Doping in the Presence of Unwanted Defects}

Figure~\ref{fig:panel3-Shallow-doping-with-unintentional-impurities}(a) shows the formation energies of $\mathrm{Be_{B}}$ and $\mathrm{Si_{B}}$ in the presence of unwanted defects, which we found to have relatively small formation energies and are therefore, likely to form in cBN.  For all of the calculations presented in this section, we explored several of the different possible sites for the shallow dopant, assuming that the unwanted defects were already present in the structure [see Supplementary Note II for additional details]. 
In this section, from amongst different possibilities for the placement of $\mathrm{Be_{B}}$ and $\mathrm{Si_{B}}$ shallow dopants relative to the unwanted defects, we are reporting our results for the lowest energy configurations. For example, Figure~\ref{fig:panel3-Shallow-doping-with-unintentional-impurities}(b) shows the the lowest energy configurations for the two shallow dopants in the presence of O$_\text{{N}}$ and one of the defect complexes, C$_\text{{N}}$O$_\text{{N}}$. 

 In the presence of a simple O$_\text{{N}}$ point defect, incorporation of beryllium as Be$_\text{{B}}$ is favorable in both B- and N-rich conditions, even more so than it was in pristine crystal [cf. Fig.~\ref{fig:panel3-Shallow-doping-with-unintentional-impurities}(a) and Fig.~\ref{fig:Shallow_Dopants}(a)]. This is due to the electrostatic attraction between the n-type defect, O$_\text{{N}}$, and the p-type defect Be$_\text{{B}}$.  On the other hand, the formation of Si$_\text{{B}}$ in the presence of O$_\text{{N}}$ requires energies very similar to those seen in Fig.~\ref{fig:Shallow_Dopants}(a).  We, also find that for all other defects as well, incorporation of Be$_\text{{B}}$ requires less energy than Si$_\text{{B}}$. 

Since the presence of the unwanted defects changes the electronic structure properties of cBN, it is important to understand how different unwanted defects modulate the desired electronic structure properties of the intentionally-doped cBN.  Fig.~\ref{fig:panel3-Shallow-doping-with-unintentional-impurities}(c) shows the projected DOS for Be$_\text{{B}}$ (top and middle panels) and Si$_\text{{B}}$ (bottom panel) in the presence of an O$_\text{{N}}$-defect in the cBN. As an O$_\text{{N}}$-defect compensates for a Be$_\text{{B}}$-defect, the structure is insulating when the concentrations of the two defects are the same [see the top panel in Fig.~\ref{fig:panel3-Shallow-doping-with-unintentional-impurities}(c)]. The inset in the top panel of Figure ~\ref{fig:panel3-Shallow-doping-with-unintentional-impurities}(c) shows the isosurface plot of a shallow state (at the $\Gamma$-point) at the valence band edge. In turn, the presence of these bulk-like shallow states at the top of the valence band implies that a simple charge-injection through gating to reposition the Fermi level in the valence band may fix the issue. Nevertheless, the incorporation of Be$_\text{{B}}$ (at this level of doping) by itself does not result in the desired ionizable shallow acceptor states. To remedy this, Be$_\text{{B}}$ will have to be doped beyond compensation limits. In order to explore this possibility, an additional Be$_\text{{B}}$ defect was created, doubling the number of Be-dopant. The PDOS for cBN doped with an additional Be-atom is plotted in the middle panel of Fig.~\ref{fig:panel3-Shallow-doping-with-unintentional-impurities}(c), showing shallow acceptor states just above Fermi level. The inset in Fig.~\ref{fig:panel3-Shallow-doping-with-unintentional-impurities}(c) [middle panel] is a isosurface plot of this shallow state. The outcome for silicon-doping (as Si$_\text{{B}}$) in the presence of O$_\text{{N}}$-defect in the cBN is shown in the bottom panel in Fig.~\ref{fig:panel3-Shallow-doping-with-unintentional-impurities}(c).  In the presence of the O$_\text{{N}}$-defect, Si$_\text{{B}}$ no longer results in shallow donor states, which can yield mobile free carriers. Instead, we find that the structure is insulating, with a very sharp (hence spatially-localized) completely filled defect state more than 1\,eV below Fermi level.  We find that this is a result of large structural distortion around and in the lattice between the two defects. The inset in the bottom panel in Fig.~\ref{fig:panel3-Shallow-doping-with-unintentional-impurities}(c) shows the isosurface plot for this state (at the $\Gamma$-point), showing its polaronic nature. This polaron is mostly localized in the region between the silicon atom, it's neighboring nitrogen atom, and a boron atom, which lies along the diagonal between oxygen atom in the middle of the cell, and the silicon atom further away along the diagonal. As O$_\text{{N}}$ is formed in abundance naturally, our results show that its presence is a major bottleneck to both n- and p-doping cBN.

 
The other defect explored here is the C$_\text{{N}}$O$_\text{{N}}$ complex, which is expected to form in abundance in cBN.  In its most favored position, the p-type beryllium-dopant prefers to substitute a boron atom next to the n-type oxygen-substituent within the C$_\text{{N}}$O$_\text{{N}}$-defect complex [see Fig.~\ref{fig:panel3-Shallow-doping-with-unintentional-impurities}(b)].  This results in a reduction in the formation energy of Be$_\text{{B}}$ in the presence of the C$_\text{{N}}$O$_\text{{N}}$-defect complex as compared to its formation energy in pristine cBN.  The ground state of the system with Be$_\text{{B}}$ in the presence of the C$_\text{{N}}$O$_\text{{N}}$-defect complex is spin polarized (net magnetic moment = 1\,$\mu_{B}$) as can be seen in the top panel of Fig.~\ref{fig:panel3-Shallow-doping-with-unintentional-impurities}(d). In addition, the PDOS plot shows that the bonding between the Be- and O-atoms lowers the energy of beryllium's $2p$-derived states, and they now lie below the Fermi level.  The states around the Fermi level are mostly derived from the carbon atom's $2p$-derived states, with some contributions from the $2p$-orbitals of the surrounding nitrogens. The inset in Figure~\ref{fig:panel3-Shallow-doping-with-unintentional-impurities}(d) [top panel] is a charge-density plot of the state just above the Fermi level, showing the localized nature of the defect state in real-space.  In contrast to the case of beryllium, doping cBN with silicon in the presence of the C$_\text{{N}}$O$_\text{{N}}$ complex results in an introduction of three defect states just below the Fermi level, two of which are very dispersive.  Fig.~\ref{fig:panel3-Shallow-doping-with-unintentional-impurities}(d) [bottom panel] is a plot of the PDOS for different impurity atoms (Si, C and O), along with the neighboring boron- and nitrogen-atoms. It shows that in the presence of the C$_\text{{N}}$O$_\text{{N}}$ defect complex, the Si $3p$ contributes to the defect states just below the Fermi level. The two charge density plots in Figure~\ref{fig:panel3-Shallow-doping-with-unintentional-impurities}(d) [bottom panel] show the nature of two of these states; only one of the dispersive states is shown.  The deeper defect state is more localized and is reminiscent of the defect state introduced by O$_\text{{N}}$, hybridized with silicon-introduced defect states.  The two shallower defect states indicate that even in the presence of the unwanted C$_\text{{N}}$O$_\text{{N}}$ defect complex, silicon doping might result in ionizable states.  We also find very similar behavior for Be- and Si-doped cBN in the presence of the C$_\text{{B}}$C$_\text{{N}}$ defect (see Supplementary Note III). Hence, with the exception of Si, which still introduces shallow states near the Fermi level in the presence of the two defect complexes, all unwanted defects studied in this work are detrimental to shallow doping of cBN.

\section{Conclusion}

This work demonstrates how the presence of unwanted defects influences the shallow doping of cBN with Si- and Be-atoms for n-type and p-type conductivities, respectively. 
Although the isolated carbon and oxygen substitutionals have been studied in the past~\cite{Orellana2000}, we find that the defect complexes, such as C$ _\text{{N}}$O$ _\text{{N}}$ and C$ _\text{{B}}$C$ _\text{{N}}$, are energetically favorable as compared to the corresponding isolated point defects, and would most likely form in cBN upon annealing. These defect complexes and other common defects (e.g. simple O$_\text{{N}}$-defect) that are inadvertently present, have a profound influence on any attempts to shallow dope cBN with Si- or Be-atoms. We find that in several cases, these unwanted defects result in deep defect states, which originate from the localized $2p$ states of the $2^{nd}$ row elements (C$_\text{N}$), and/or result in removal of desirable shallow states that are needed to create mobile free carriers. These findings are an important step towards understanding the role of unintentional defects in the observed doping-limitations of cBN. In turn, it will allow us to find solutions to the issues created by unwanted defects that will allow controllable n- and p-doping of cBN for extreme applications. These protocols may include determining the conditions under which the formation of unwanted defects is suppressed and/or finding doping levels that go beyond compensation limits. Concrete examples of such defect-limiting protocols may include: (i) limiting formation of O$_\text{{N}}$ in cBN, and (ii) limiting formation of both C$_\text{{N}}$O$_\text{{N}}$ and C$_\text{{B}}$C$_\text{{N}}$-defect complexes by minimizing carbon-based contaminants. In addition, under the the N-rich conditions, not only the the formation of C$_\text{{N}}$ and C$_\text{{N}}$-based complexes would be limited, but intentional doping with Be and Si is also easier.
\section{Acknowledgement}
This work is supported by National Science Foundation (NSF Grant Nos ECCS-1831954, PHY-1659224 and DMR-1231319). We acknowledge the computational support provided by the Extreme Science and Engineering Discovery Environment (XSEDE) ~\cite{xsede,xsede-ecss} under Project PHY180014, which is supported by National Science Foundation Grant No. ACI-1548562. For three dimensional visualization of crystals and volumetric data, use of VESTA 3 software is acknowledged.

\normalem      


\end{document}